# Private Information Disclosure from Web Searches

## (The case of Google Web History)

Claude Castelluccia[1], Emiliano De Cristofaro[2], Daniele Perito[1]

[1] INRIA Rhone Alpes, Montbonnot, France
{claude.castelluccia, daniele.perito}@inrialpes.fr

[2] Information and Computer Science, University of California, Irvine
edecrist@uci.edu

**Abstract.** As the amount of personal information stored at remote service providers increases, so does the danger of data theft. When connections to remote services are made in the clear and authenticated sessions are kept using HTTP cookies, data theft becomes extremely easy to achieve. In this paper, we study the architecture of the world's largest service provider, i.e., Google. First, with the exception of a few services that can only be accessed over HTTPS (e.g., Gmail), we find that many Google services are still vulnerable to simple session hijacking. Next, we present the Historiographer, a novel attack that reconstructs the web search history of Google users, i.e., Google's Web History, even though such a service is supposedly protected from session hijacking by a stricter access control policy. The Historiographer uses a reconstruction technique inferring search history from the personalized suggestions fed by the Google search engine. We validate our technique through experiments conducted over real network traffic and discuss possible countermeasures. Our attacks are general and not only specific to Google, and highlight privacy concerns of mixed architectures using both secure and insecure connections.

> **Update:** Our report was sent to Google on February 23rd, 2010. Google is investigating the problem and has decided to temporarily suspend search suggestions from Search History. Furthermore, Google Web History page is now offered over HTTPS only. Updated information about this project is available at:
> `http://planete.inrialpes.fr/projects/private-information-disclosure-from-web-searches`

## 1 Introduction

With the emergence of cloud-based computing, users store an increasing amount of information at remote service providers. User profiling techniques can complement such information automatically. Cloud-based services often come at no cost for the users, while service providers leverage considerable amounts of user profiling information to deliver targeted advertisement. Such a business model appears to benefit all parties. However, storing large amounts of personal information to external providers raises privacy concerns. Privacy advocates have highlighted the conceptual and practical dangers of personal data exposure over the Internet [12,13,14,15].

In this paper, we analyze private information potentially leaked from web searches to third parties, rather than focusing on data disclosed to service providers.

**The case of Google Web History.** Being the world's largest service provider (according to alexa.com), we focus on the case of Google. In particular, we analyze one Google service: Web History. It provides users with personalized search results based on the history of their searches and navigation. Such a history is accessible at `http://google.com/history`. For more details, we refer to Section 2.

Web searches have been shown to be often sensitive [15]. Any information leaked from search histories could endanger user privacy. For example, the spread of influenza and the number of related search queries divided by region has been successfully correlated [17]: this suggests that search histories contain health-related data and possibly other personal information, including, but not restricted to: political or religious views, sexual orientation, etc. Furthermore, AOL's release in 2006 of 20 million nominally anonymized searches underlined that search queries contain private information [11].

The privacy of personal data stored by service providers has been long threatened by the well-known attacks consisting of hijacking user's HTTP cookies[1]. These attacks have been addressed by Google in several ways. For instance, "sensitive" services such as Gmail now enforce secure HTTPS communication by default and transmit authentication cookies only over encrypted connections. As for the privacy of the Google Web History, its login page states: "To help protect your privacy, we'll sometimes ask you to verify your password even though you're already signed in. This may happen more frequently for services like Web History which involves your personal information". Frequently requesting users to re-enter their credentials can thwart the session hijacking attack, however, as illustrated in this paper, such an attack can still be effective if a user has just signed in. Moreover, we show that search histories can still be reconstructed even though the Web History page is inaccessible by hijacking cookies.

**The Historiographer.** To this end, we successfully design the Historiographer, an attack that reconstructs the history of web searches conducted by users on Google. The Historiographer uses the fact that users signed in any Google service receive suggestions for their search queries based on previously-searched keywords. Since Google Web Search transmits authentication cookies in clear, the Historiographer—monitoring the network—can capture such a cookie and exploit the search suggestions to reconstruct a user's search history. We refer to Section 3 for more details on the reconstruction technique.
Note that such an attack is much more powerful than constantly eavesdropping on unencrypted user navigation: within a few seconds of eavesdropping, it can reconstruct a significant portion of a user's search history. This may have been populated over several months and from many different locations, including those from where a privacy-conscious user avoids sensitive queries fearing traffic monitoring. Also, the Historiographer is *non-destructive*, i.e., it does not affect user data.

Although the Historiographer builds on features and technicalities specific to the Google architecture, our goal is not to attack Google nor any particular service provider. Instead, we highlight the general problem of protecting user privacy when using a mixed architecture drawing from personal data with both secure and insecure connections.

**Contributions.** This paper makes the following contributions:

1. We show that the Google infrastructure is vulnerable to the Historiographer, a new attack that reconstructs part of the search history of users.
2. We show that the well known session hijacking attack is still applicable to many Google services. More specifically, we evaluate the security of several Google services, including Web History, against this simple attack and report the number of services vulnerable along with the amount and type of information potentially disclosed by each service.
3. We conduct an experimental analysis over network traces from a research institution, a Tor [1] exit node, and the 20 million anonymized searches released by AOL in 2006, in order to assess the number of potential victims and the accuracy of our attack[2]. Results show that almost one third of monitored users were signed in their Google accounts and, among them, a half had Web History enabled, thus being vulnerable to our attack. Finally, we show that data from several other Google services can be collected with a simple session hijacking attack.

**Paper Organization.** The rest of the paper is organized as follows. Section 2 presents the necessary technical background. Section 3 details the new Historiographer attack. Section 4 describes our experimental evaluations on real network traffic, and estimates the number of potential victims and the accuracy of the Historiographer. Independently of Historiographer, this section also evaluates the additional information leaked from Google's services through simple session hijacking. Section 5 discusses possible countermeasures to thwart the Historiographer attack, while Section 6 overviews related work. Finally, Section 7 concludes the paper.

---
[1] In a session hijacking attack, an attacker monitoring the network captures an authentication cookie and impersonates a user. In Section 6, we will discuss several vulnerabilities and concerns involving Google [8,18,24].

[2] Another similar experiment is currently in progress at a large University campus, and we will add the results in the final version of the paper.

## 2 Background

In the following, we present background information on several aspects discussed throughout the rest of the paper: the HTTP cookies, and the Google architecture.

### 2.1 HTTP Cookies

The need of maintaining sessions in HTTP emerged with the creation of the first web applications (e.g., e-commerce websites), as HTTP is a stateless protocol. RFC2109 [21] and RFC2965 [22] specified a standard way to create stateful sessions with HTTP requests and responses. They describe two new headers, Cookie and Set-Cookie, which carry state information between participating origin servers and user agents. A Cookie, which contains a unique identifier, is typically used to store user preferences or to store an authentication token.

Cookies are set by the server as follows. After an incoming HTTP request, a server sends back a HTTP response containing an HTTP header, referred to as Set-Cookie, requesting the browser to store one or several cookies. Such a header is in the form of *name=value*, the so-called "cookie crumb". As a result, provided that the user agent enables cookies, every subsequent HTTP request to a server on the same domain will include the cookie in the Cookie HTTP header. A cookie may also include an expiration date[3], or a flag to mark it *secure*. In the latter case, the browser will send the secure cookie only over encrypted channels, such as SSL.

A set-cookie header may optionally contain a *domain* attribute, which specifies the domain validity of the cookie. If this attribute is set, the cookie is referred to as domain cookie, as opposed to host cookie which is not specific to any particular sub-domain. For example, as we will present in Table 1, a user accessing Google's Calendar receives a domain cookie for `calendar.google.com` as an authentication token. Such a cookie is then to be included in every subsequent HTTP requests to the domain. In contrast, other Google's applications (such as the Search, History or Maps) only set host cookies, which are used across different services and domains.

Finally, a set-cookie header may specify a path attribute to identify the subset of URLs for the cookie's validity. For example, as we will present in Table 2a, a user that signs in Google receives three cookies, namely `SID`, `SSID`, and `LSID`. While the latter only applies to the path "*/account*", the other two are can be used for different paths.

### 2.2 Google Architecture

As we mention in Section 1, we focus on the case of the world's largest service provider, i.e., Google. This section describes the Google architecture[4].

**Google Web Products.** Google offers more than 40 free Web services, including several search engines (e.g. Google Web Search), maps (Google Maps), as well as personalized subscription-based services like email (Gmail), documents (Google Docs), photos (Picasa), videos (Youtube), Web history.

Even though some services can be used without registration (e.g., search), other are user-specific (e.g. Gmail) and require user authentication. Most of the services can be used by means of a single Google account, a combination of username and password. However, services that do not mandate registration provide extra features if users are signed in. For instance, an authenticated user can obtain personalized, potentially more accurate, search results on Google Maps based on her default location.

---
[3]If an expiration date is provided, cookies survive across browser sessions, and are then called "persistent". Otherwise, the cookie is deleted when closing the browser.

[4]Since not all the components of the Google architecture are public, some of the details presented in this section might not completly accurate.

**Google Web History.** This *opt-out* service – previously known as Google Search History and Personalized Search – is implemented by Google to provide signed-in users with personalized search results based on the history of their searches and navigation. Furthermore, users typing search queries in the Web interface are prompted with *suggestions* resulting from their history. To this end, Google tracks all Web searches performed by a signed-in user (with Web History service enabled), as well as the target web pages clicked from the search result page. This service may be further enhanced by installing the Google Toolbar, allowing Google to also track *all* visited web sites, independently from the use of the search engine. Google Web History also provides a Web interface at `google.com/history`, allowing users to view and delete their history. Users are given the choice to *pause* Web History by accessing their account. Nevertheless, Google customizes searches and provides suggestions based on data recorded before pause.

Note that Google is offering Personalized Search not only to signed-in but also to signed-out users. In fact, for these users Google performs the customization using the information linked to the user's browser with the help of an "anonymous" cookie. Specifically, Google stores up to 180 days of activity linked to such cookie. Again, users can explicitly disable this feature [3].

**Google Authentication.** Google services are accessible with a single set of credentials, composed by a pair username/password. Different services are usually hosted as sub-domains of `google.com` (or other Top-Level Domains for different countries) and offer seamless integration between each other to minimize the need for users to re-enter their credentials. Integration is achieved through the Accounts service. In practice, requests to authenticate to a Google service are redirected to the Accounts page where the user is asked to enter her username and password. If authentication succeeds, a browser cookie is set (or refreshed) to track the session and the user is redirected back to the page that was originally requested. An illustration of this mechanism is provided in Fig. 1.

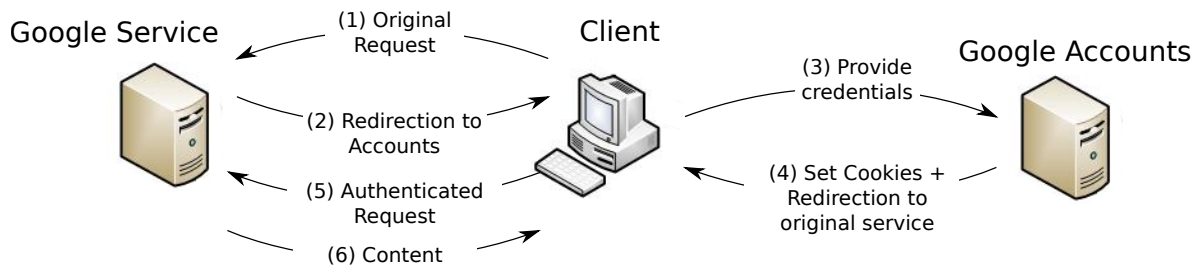

Fig. 1: The Google Accounts authentication management for Google services.

Access to Google Accounts is always secured using HTTPS. However, subsequent connections might revert back to simple HTTP depending on the requested service. For example connection to Maps Search are established with HTTP whereas HTTPS access to Gmail is enforced.

Table 1 compares several Google Services. It may be the case that services considered more sensitive are protected by HTTPS, whereas those judged less sensitive are left unencrypted. In particular, we noticed that the use of HTTPS is mandatory for some services (e.g., Gmail), while impossible for others (e.g., Search). Additionally, there are services accessed on HTTP by default, but users may force a secure connection specifying `https://` in the URL.

**Google cookies.** Authenticated sessions are kept by means of cookies that are set by Accounts upon successful authentication. Two cookies, called `SID` and `SSID`, are used as authentication tokens across most services[5] for unencrypted and encrypted connections, respectively. We believe their names might stand for *Session ID* and *Secure Session ID*[6].

| Service Name | Default Connect. | HTTPS Support | Domain specific cookie | Purpose |
|---|---|---|---|---|
| Search | HTTP | no | no | Web search |
| Maps | HTTP | no | no | Maps search |
| Reader | HTTP | yes | no | RSS/Atom feed reader |
| Contacts | HTTP | yes | no | Address book manager |
| History | HTTP | yes | no | Search history manager |
| Gmail | HTTPS | mand. | no | Web mail application |
| Accounts | HTTPS | mand. | no | Google account manager |
| News | HTTP | no | no | News aggregator |
| Bookmarks | HTTP | yes | no | Bookmark manager |
| Docs | HTTP | yes | yes | Office application |
| Calendar | HTTP | yes | yes | Calendar application |
| Groups | HTTP | yes | yes | Discussion groups application |
| Books | HTTP | no | no | Personalized digital library |

Table 1: Some of Google's services

A description of several Google cookies is reflected in Tables 2a and 2b. Note that: (1) `SID` and `SSID` are valid for all Google sub-domains and are used to authenticate users to several services, (2) `SID` is not a secure cookie, i.e., it is sent on every connection to Google, while `SSID` is only sent over encrypted connections, and (3) `NID` represents the "anonymous" cookie used to track unlogged users. There are also a number of cookies not reported, which are used for miscellaneous purposes, e.g., to store language or search interface preferences.

In our study, we will focus on the `SID` cookie, providing authenticated access to most unencrypted services. In particular the `SID` cookie is set in all web searches. It is used by Google to identify the requesting account, populate the account's Web History and provide personalized web results and suggestions.

## 3 Historiographer: Reconstructing Search History

### 3.1 Attack Overview

In the following, we present the Historiographer, an attack aiming to reconstruct users' search histories stored by Google. The attack consists of two steps.

First, it hijacks a session stealing the victim's `SID` cookie. This can be done, for example, by eavesdropping on her traffic, and in particular on any request to a Google service, such as Google search. Eavesdropping

---

[5] All services that do not use domain cookies, such as Maps, History, Search, Reader, Books and Contacts – see Table 1.

[6] An additional list of domain-specific cookies, such as those for `docs.google.com` or `calendar.google.com`, are sent in clear text but are *set* only over a secure connection upon user access.

| Cookie-Name | Secure | Domain | Path | Purpose |
|---|---|---|---|---|
| SID | no | google.com | / | authentication token |
| SSID | yes | google.com | / | secure authentication token |
| LSID | yes | google.com | /accounts | secure authentication token |

(a) Google's cookies for signed in users

| Cookie-Name | Secure | Domain | Path | Purpose |
|---|---|---|---|---|
| NID | no | google.com | / | track unlogged users |
| PREF | no | google.com | / | store search settings (e.g., language) |

(b) Google's cookies for not signed in users

Table 2: Description of the type and purposes of some cookies used in the Google platform.

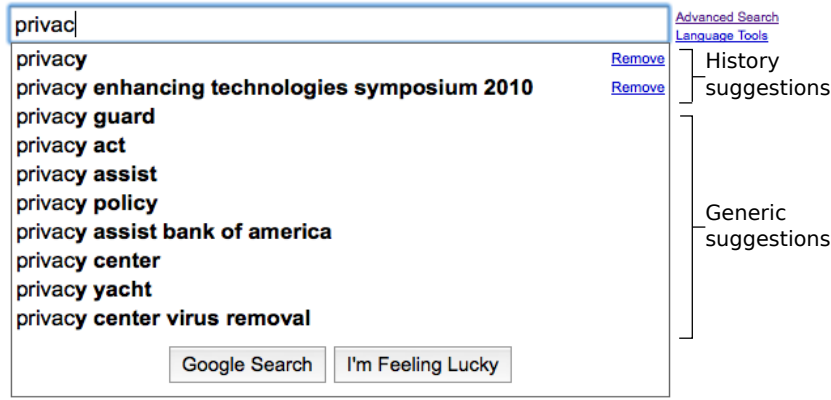

Fig. 2: An example of Google Search Suggestions.

can be performed by listening on a local wired network, an open wireless network, such as a campus network, or by deploying a Tor exit node (as detailed in Section 4). This does not necessary involve compromising nodes, and therefore does not require special skills.

Second, it reconstructs the Web History using as *partial precise* inference [16]. We recall that an inference attack is a technique used to disclose sensitive and protected information from presumably non-sensitive data. In this setting, we reconstruct part of the potentially privacy-sensitive Web History from web searches. The technique is *partial*, because, as shown in Section 3.2, it does not always reconstruct the whole history. Finally, it is *precise* since it infers accurate items from the Web History without introducing errors, as opposed to imprecise inference techniques that do it with a certain probability.

Note that any user, in particular if equipped with a mobile device, is likely to access the Internet via an unencrypted wireless channel at some point of time. As soon as she signs in Google when connected to such unprotected networks, she becomes vulnerable to our attack. Furthermore, the attack is effective even if the user is careful and never inputs sensitive information during "insecure" browsing sessions over unencrypted wireless channels.

### 3.2 Reconstructing the web history through inference

**Web history access control.** Authenticated users can consult, modify, pause or delete their complete history by accessing the Web History service. The history can be consulted as an HTML page or an RSS feed. However, as mentioned above, Google access control policy for Web History differs from the one implemented in other services. In fact, users are frequently asked to re-enter their credentials even though they are already authenticated. Preliminary tests showed that this mechanism is used quite frequently, and in such a case a session hijacker would be prevented from downloading the history.

**Exploiting the search suggestion feature.** However, a feature provided by Google, namely the *search suggestions*, helped us circumvent the access control enforced for Web History. As mentioned in Section 2, Google search engine offers contextual information in the search interface that can be derived from the user's search history. Specifically, whenever a prefix is typed in the search box, an Ajax [20] request is sent to a Google server, which replies with a list of associated keywords. Fig. 2 presents an example of a user typing the prefix "privac" in the search box. The user is then prompted with a list of related keywords to auto-complete the search, i.e., *search suggestions*. These keywords can either be based: (1) on Google's ranking of similarity (we call them *generic* search suggestions), or (2) on user's search history (we call them *history* search suggestions). Note that history search suggestions are only sent to the user if the typed prefix corresponds to search queries that are in the Web History and were followed by a "click" on one of the results. We call these queries: "clicked" queries. History search suggestions are visually distinguishable from the generic ones, since they include a link to remove them. This is reflected in the Javascript code, as history search suggestions have a flag set to differentiate them. The access to the web server that implements suggestions is carried out using Ajax and every request is authenticated sending an SID cookie, which can be easily eavesdropped and hijacked.

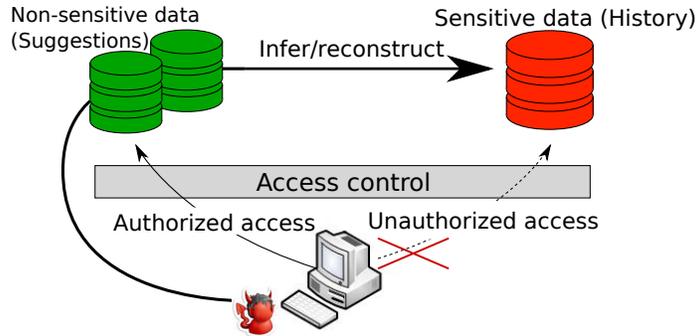

Fig. 3: Reconstructing history. Since the attacker is prevented from accessing the History service, she uses non-sensitive data from suggestions to infer the sensitive search history.

Therefore, once an SID cookie has been captured, the user's Web History can be reconstructed using the suggestion service: Historiographer steals a user authentication cookie and then sequentially requests possible prefixes to the suggestion server to recover keywords coming from the history. An illustration of the general reconstruction attack can be seen in Fig. 3.

**Reconstruction algorithms.** In order to reconstruct a search history, the Historiographer needs to ask for suggestions for different prefixes. Hence, we need to carefully select the list of possible prefixes to use, since the keywords in the history are unknown. We encounter the following obstacles: (1) the number of requests for suggestions should be kept to a minimum, in order to be as stealthy as possible; (2) at most three replies come from the suggestion history upon each suggestion request, limiting the amount of information discovered with each request; and (3) suggestions are only returned for two-letter (or longer) prefixes, preventing from simply looking for all letters in the given alphabet.

A naïve (*brute-force*) approach would involve requesting all the possible two- or three-letter prefixes to harvest the replies coming from the history. However, this would already require $26^2 = 676$ requests for two-letter and $26^3 = 17,576$ for three-letter combinations in the English alphabet, hence relatively high numbers that might lead to detection.

Instead, the Historiographer employs a more sophisticated technique: it requests only prefixes that are common in a given language. For instance, if one considers English, there are only 7 words starting with the two-letter prefix oo, while no word starts with the prefix qr. Whereas, the most used prefix results to be co, used in 3223 words. It is then reasonable to expect that in the search history there are more entries starting with the letters co than with qr.

As a result, we proceed as follows: We extract all two-letter prefixes from a reference corpus, order them by frequency, and we select only the prefixes in the $90th$ percentile. We used two different reference corpora in our experiments: the English dictionary and the AOL dataset of 20 million anonymized searches that was released in 2006 [11]. However, they both achieved very similar performance. For the English dictionary, this yields a total of 121 two-letter combinations and reduces the number of requests and the fingerprint of the attack[7].

Further, we notice that at most 3 search suggestions can come from the history for each requested prefix. Thus, if we get exactly 3 suggestions from the history, there are either 3 or more search queries starting with the corresponding prefix. This is a potential indication that this prefix is particularly frequent in the history, and it is worth being further explored. Hence, whenever we encounter a two-letter prefix producing 3 suggestions, we add another letter to the prefix and we repeat the request. Note that the resulting three-letter prefix is again generated by extracting the most common three-letter prefixes from the dictionary and not by simply adding every possible letter in the alphabet. Fig. 4 visually depicts this procedure: The prefix `co` produces 3 results and is further explored, contrary to `de` and `ya` who produce only 2 (resp., 1) results. A description of the achieved accuracy and the related overhead in terms of requests is provided in Section 4.2.

---

[7]Different languages can be supported by simply changing the alphabet and the reference corpus.

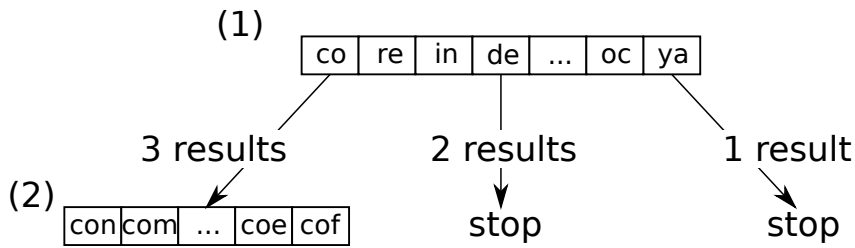

Fig. 4: Smart tree approach. To reconstruct large portions of the search history, we start with the most common two-letter prefixes (1). If a prefix produces 3 suggestions, then we descend in the tree (2).

**Implementation.** We implemented the Historiographer as a Perl application. It is part of a more complete tool that: (i) captures traffic from a network interface, (ii) recognizes cookies sent to and from Google servers, and (iii) then uses them to hijack sessions and retrieve personal information. Web History is only one of the services the software collects data from.

### 3.3 Beyond the Historiographer: Exploiting Personalized Results

The Historiographer attack uses Personalized Search to leak information from a user's Web History. However, one could also use the so-called Personalized Results, i.e., the fact that search queries on Google often produce different results based on the user's search history. We present an example of this in Fig. 5. If the results contain at least one linked page previously accessed by the user, the "View customizations" link appears at the top right corner of the result page. One can easily identify the visited linked pages (e.g., http://petsymposium.org/2010/ in Fig. 5) since they are marked with a tag reporting the number of visits (e.g., 8), and the date of the last visit (e.g., March 1st).

Therefore, an adversary can verify that specific keywords belonging to a user's search history using the Personalized Results. We call such an attack a *targeted check*. Note that the adversary does not have to test the exact matching keyword searched by the user. It is enough to make a related search that includes the visited linked page in the results. For instance, assuming that a user has searched for *PETS 2010* and *Oakland 2010*, and has then clicked on the related links http://petsymposium.org/2010/ and http://oakland09.cs.virginia.edu/. A subsequent search for the keyword *Privacy* would produce a result page with the "View customizations" link. Looking at the result page produced by only one request, an adversary can find out that the above pages were visited and conclude that the user is interested in *privacy*, in *PETS 2010* and *IEEE Security and Privacy*. The adversary could then try other keywords and broaden the information leakage or profile user's interests. Note that this attack can be amplified with the exploitation of the new Google's **Star** service that allows users to mark their favorite web sites. With **stars**, a user can mark his favorite sites by simply clicking the star marker on any search result or map. As a result of this action, these sites will appear in a special list next time the identical or a related search is performed. This feature gives even more power to the adversary.

Note that this attack only applies to signed-in Google users with Web History enabled (a significant proportion of Google users as showed in Sec. 4). However, as discussed in Sec. 2, Google provides customized searches to signed-out users too, using an "anonymous" cookie. Therefore, we believe that a similar attack can be designed for signed-out users as well, although the history would be limited to the life of this cookie.

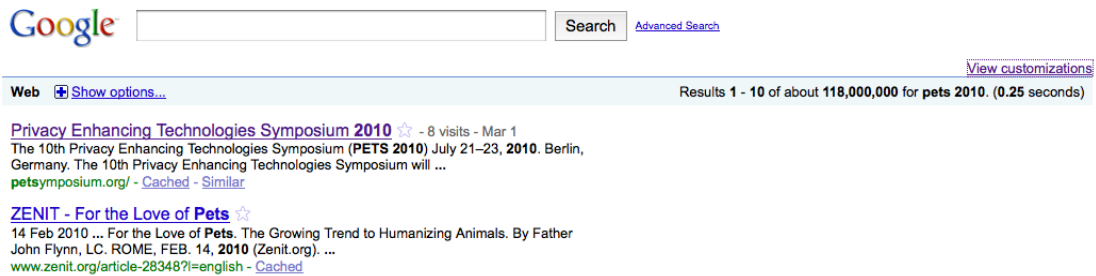

Fig. 5: An example of Google Personalized Results.

## 4 Measurements and Analysis

Given the private nature of the information gathered and the difficulty of having users willing to disclose them, we conducted three different experiments. These experiments were aimed at collecting data to estimate: (1) the number of potential victims that access Google services while being signed-in and, among them, how many have Web History enabled; (2) the accuracy and the cost of the Historiographer; (3) the amount of private data that can be retrieved from other services with the simple session hijacking attack.

In addition we issued a survey to 68 users to estimate whether they consider the information stored in search histories to hold private and sensitive data. We refer to Appendix B for more information on the survey.

### 4.1 Estimating the number of potential victims

In order to estimate the number of potential victims of our attack, we conducted an experimental analysis on the network traces collected from a research center with about 500-600 daily users (more details will be provided after the review process is completed) and a Tor exit node. We collected one week of network traffic during February 2010. The goal was to measure the percentage of users using Google while signed-in, and that having the Web History service enabled. Note that only aggregate data was stored. The data collected from the research center was analyzed passively, i.e., no session was actually hijacked.

In order to count the number of users from a network trace, one needs reliable identifiers to filter out duplicate queries or changes of network identifiers, e.g. IP churn. Luckily we could use cookies gathered from the network captures to identify single users. As explained above, Google issues persistent cookies both to signed-in and not signed-in users. Among them we chose to use `SID` cookies to identify signed-in users and `NID` cookies to identify not signed-in ones. Furthermore, in order to count the number of users with history enabled, our application looked for a particular link to the History service that is included in each search result page.

The results of test are presented in Table 3a. Around one third of the users resulted to be signed-in while using Google services, including web searches. Furthermore, about half of the users with an account have history enabled. The limited size and the lack of randomness in the choice of our sample, does not allow us to draw conclusions about the entire population of users. However, if we combine our results with the above mentioned popularity of Google services, it would appear that a significant portion of web users are at risk.

### 4.2 Estimating Historiographer's accuracy

**Volunteers.** In order to evaluate the extent of potential leakage of private information from Google web searches, we turned to volunteers. It would have been otherwise impossible to conduct our study on uninformed users without incurring legal and ethical issues. We aimed at evaluating the accuracy of the Historiographer at reconstructing web histories. To this end, we "attacked" the accounts of 10 volunteers using our software and measured its accuracy.

| Experiment | Number of Google users | Number of users signed-in | Number of signed-in users with History enabled |
|---|---|---|---|
| Research center | 1502 | 543 (36.1%) | 223 (14.8%) |
| Tor Exit Node | 1893 | 872 (46.1%) | 441 (23.29%) |

(a) Measurements on network traces

| User ID | $n_h$ | $n_c$ | $n_s$ | Recall | $n_{requests}$ | History Activation date |
|---|---|---|---|---|---|---|
| 1 | 751 | 442 | 308 | 0.69 | 680 | Aug 08, 2009 |
| 2 | 318 | 142 | 99 | 0.69 | 368 | Mar 10, 2008 |
| 3 | 621 | 321 | 176 | 0.54 | 483 | May 16, 2009 |
| 4 | 520 | 248 | 169 | 0.68 | 400 | May 22, 2007 |
| 5 | 657 | 309 | 231 | 0.75 | 601 | Feb 06, 2009 |
| 6 | 389 | 202 | 130 | 0.64 | 365 | Fen 12, 2009 |
| 7 | 690 | 337 | 201 | 0.60 | 560 | Jul 18, 2008 |
| 8 | 416 | 219 | 143 | 0.65 | 399 | Aug 09, 2006 |
| 9 | 228 | 127 | 69 | 0.54 | 211 | Aug 20, 2008 |
| 10 | 306 | 164 | 118 | 0.72 | 334 | Sep 27, 2009 |
| 11 | 1567 | 930 | 506 | 0.54 | 740 | Oct 26, 2009 |
| 12 | 1163 | 680 | 533 | 0.78 | 823 | Dec 4, 2009 |

(b) Results from volunteers

| Type of information leaked | Corresponding service | Number of Accounts accessible | Mean number of entries collected |
|---|---|---|---|
| Complete (unrestricted) Search History | History | 45(5%) | 123 |
| Blogs followed on Reader | Reader | 139(15%) | 14 |
| Address book | Contacts | 766(87%) | 189 |
| Maps search history | Maps | 696(79%) | 22 |
| Default address on Maps | Maps | 52(5%) | 1 |
| Financial portfolio | Portfolio | 11(1%) | 8 |
| First/Last name | Maps profile | 661(75%) | 1 |
| Bookmarks | Bookmarks | 236(27%) | 79 |

(c) Aggregate Information analyzed from 872 Tor users.
Table 3: Results from the three experiments.

The performance of the Historiographer at reconstructing search histories can be measured in terms of *recall*. For every user $u$, we call $H$ the set of entries in $u$'s history, $H_c$ the subset of searches whose results were clicked by $u$, and $S$ the set of entries reconstructed from suggestions. We denote $n_h = |H|$, $n_c = |H_c|$ and $n_s = |S|$. Since suggestions are only given for "clicked" queries, the recall $R$ of our reconstruction algorithm can be measured as the ratio $R = \frac{n_s}{n_c}$.

Results are reflected in Table 3b. The Historiographer reconstructs a significant portion of a user's history, with a mean recall of 0.65. The mean number of requests per user to reconstruct the history was 440. Since users are kept signed-in for two weeks, these requests can be made at a low pace to increase stealthiness. For instance, an attacker could issue a request every hour and still expect to retrieve 65% of the "clicked" queries. Also, the recall can arbitrarily be increased by increasing the number of requests. On average, with about 2000 requests, we can obtain a mean recall of 0.81. The mean recall lowers to 0.34 when considering the ratio of reconstructed entries over the complete set $H$.

Recall that the Historiographer can only recover "clicked" queries, although, as we illustrate in Appendix A, a complete history typically contains more information and additionally stores the time and the frequency of searches. We argue that only recovering "clicked" queries is not a tremendous limitation. When inspecting volunteers' history, we noticed that "clicked" queries are often corrections of generic or misspelled queries. A more accurate analysis of this phenomenon is left for future work. Note also that the Google's algorithm producing keyword suggestions is based on several parameters, such as dates and frequencies of searches and visited web sites. Therefore, we believe that the accuracy and the amount of information that can be retrieved by the Historiographer could be further improved with a deeper understanding of the underlying algorithm. On the other hand, it appears that the likelihood that an entry in the history is returned as a suggestion decreases over time, which could negatively affect the recall for older entries.

**AOL Dataset.** Next, we tested our attack on a wider sample. We used the anonymized query dataset released by AOL in 2006, containing 20 million searched made by 650,000 users. From the dataset, we constructed the search history of each user. Then, simulating the search suggestions fed by Google drawing from the histories, we estimated the recall of our reconstruction technique. The mean recall was 0.64, an accuracy similar to that obtained for the volunteers.

### 4.3 Additional Information Leakage via Session Hijacking

As mentioned above, in addition to the Historiographer, an attacker can hijack a user's session to access several Google services. This section evaluates the extent of the information leaked. We ran our software for a week on a Tor exit node, and we analyzed 872 Google accounts. We stress that our software only generated aggregate data automatically and discarded the information immediately.

Note that we used Tor only as a way to collect anonymized network traces. This cannot, by any means, be considered as an attack against Tor. In fact, even considering a malicious Tor exit node, the attacks can be prevented by using the appropriate tool configuration to block cookies transmitted over HTTP. (For more information, we refer to [4,5]). However, we point out that a significant number of users are not aware of the dangers. In fact, they authenticate to Google while connected in Tor and do not block HTTP cookies, thus endangering their anonymity and privacy to potential malicious Tor exit nodes. In fact, a malicious entity could set up a Tor exit node to hijack cookies and reconstruct search histories. The security design underlying the Tor network guarantees that the malicious Tor exit node, although potentially able to access unencrypted traffic, is not able to learn the origin of such traffic. However, it may take the malicious node just one Google `SID` cookie to reconstruct a user's search history, the searched locations, the default location, etc., thus significantly increasing the probability of identifying a user.

Additional example applications include RIAA tracking users that ever searched—although connected into Tor—for torrent files related to unlicensed material.

**Session Hijacking Attack.** By means of session hijacking, we tried to access the following information: locations searched on Maps (along with the "default location", when available); blogs followed on Reader; full Web History (when accessible without re-entering credentials); finance portfolio; bookmarks. For each of them, we counted the number of entries retrieved and reported the mean over the 872 accounts. Table 3c summarizes the obtained results.

We point out that for 5% of the accounts, we accessed the Web History page without being asked to re-enter credentials (simply replaying the `SID` cookie). We stress that the session hijack had a significant success rate for many popular services. For instance, we retrieved 79% of the searched locations on Maps and the 87% of address books (Contacts). Also, we were able to retrieve the first and last name associated to the account in 75% of cases. Unfortunately, *these numbers translate into a significant amount of personal (and identifying) information leaked through session hijacking*.

Notably, the information collected from the Maps service was composed of maps queries coming from the histories of the users. Similarly to history suggestions, users that access Maps are presented with entries that come from the locations they previously searched for. Differently from search suggestions, Maps suggestions are not the result of an prefix based Ajax query to a remote Google server. Instead, for signed-in users, the page at `maps.google.com` includes a Javascript array that includes *a*ll previous searches. Accessing this information only requires retrieving the web page once and does not require the use of the Historiographer. A portion of this list is shown in Example 1. The provided information is very detailed and includes: the exact location searched (address:), the time, in seconds since the Epoch, it was searched (created:) and the number of times the location was searched (count:). The information collected this way is of the same kind of the one collected by the Historiographer but referred to maps searches instead of generic web searches. However, the specifics of the design of Maps suggestions make the attack on this service much easier. We can only speculate on the reasons behind such a design. One could be that, since Maps history is relatively small in mean size 3c, it is more efficient to send all the information at once, rather than relying on multiple Ajax requests and replies. Whatever the reasons, this design makes location information stored on Google more vulnerable to session hijacking than search history.

```
{id:19,address:"1600 Amphitheatre Parkway Mountain View",label:"",created:1254038860,count:13,...},
{id:20,address:"Piazza di Spagna, 00187 Roma, Italy",label:"",created:1254251745,count:2,...},
{id:21,address:"Newark, CA",label:"",created:1255123644,count:1,...}
```

Example 1: Portion of the Javascript code that includes Maps history information.

### 4.4 Web history and Smart Phones

With the increasing number of smart phones users, search history is likely to be strongly correlated with users' location of the users. We noticed that Google maintains a separated Web History when the search page is accessed from an iPhone. Such a history has a less strict access control policy. Similarly to Google Maps, the whole search history is sent as a Javascript list embedded in the page. Supposedly, this information is presented only when using the iPhone. However, one just needs to set the appropriate user agent string when accessing Google (for example through the User Agent Switcher Firefox extension [7]). Then, replaying the SID cookie, the whole Web History becomes accessible, with a single page access.

We tested this strategy on the set of volunteers. We were able to retrieve their iPhone search history from a regular PC by switching the PC's browser user agent to an iPhone user agent, and hijacking the victims' SID cookies.

## 5 Possible Countermeasures

The vulnerability targeted by the Historiographer is difficult to address because of the complexity and scale of the Google architecture, as well as the performance and usability requirements. However, we discuss some possible countermeasures. For instance, users could take the following precautions, simultaneously: (i) always log out from any Google service when performing a search, (ii) disable the Web History service, and (iii) disable personalization from anonymous cookies or always delete Google cookies, similarly to what is suggested by the Electronic Frontier Foundation On the other hand, Google could either: (i) discontinue the Personalized Search service, or (ii) let the users choose to enforce HTTPS for web searches (for instance, by clicking on a special link when surfing from *insecure* networks) and trade off speed with privacy. However, one can argue that solutions preventing personalized searches may degrade the service, whereas the use of HTTPS on Web Search[8] may be too expensive to put in place. Evidence of this is given by the impossibility of accessing Google search page via HTTPS and by the concerns already expressed by Google regarding the performance of using HTTPS for Gmail [9].

**Compartmentalized Searches.** We propose an additional mitigation technique that would allow to keep the Personalized Search service. Specifically, we propose that Google could keep separate histories based on the networks from which user's searches originate. Then, it can provide different search suggestions (and personalized results) based on different locations. We imagine an extension to the `google.com/history` web page to allow a user to configure such locations and the privacy settings related to them. Although this would not solve all possible information leakage, it would compartmentalize user's private information: Consider for instance an employee reluctant to reveal personal information to her employer (e.g., that she is looking for another job). Fearing that her navigation within the company network is monitored, she might avoid accessing potentially "compromising" information. If she signs in Google from the company network, however, her search history —containing for instance "compromising" searches made from home—(and more) can be leaked.

**Binding authentication cookies to IP addresses.** Several web sites, e.g., LiveJournal [2], allow user agents to bind the authentication cookies to the current IP address. In other words, the server does not accept an authentication cookie that originates from a different IP address. However, this technique is not always enforced due to drawbacks on the usability of the service. For example, "mobile" users, whose IP address often changes, would be forced to frequently re-enter their credentials. However, depending on the network configuration, binding cookies to IP addresses could not be enough to prevent session hijack. For instance, an attacker operating on a local network could succeed by poisoning the ARP table on the local Ethernet switch. Note also that at the moment Google allows a single account to be signed-in from multiple locations and with multiple IP addresses (although some services such as Gmail display the number of simultaneous connections at the bottom of the page).

---

[8]Note that adopting HTTPS only for the Web History web page would not prevent the Historiographer, but only the access to the page.

## 6  Related Work

To the best of our knowledge, this work is the first to focus on the private information leaked from web searches to third parties. In the following, we present the most relevant work to several concepts and tools that we use.

**Session hijacking.** Since their early appearances, the use of cookies to maintain authenticated sessions has lead the way to *session hijacking* attacks (see for instance [18]). These attacks are quite simple: an attacker monitoring network traffic may sniff an authentication cookie and replay it to impersonate another user. For this reason, sensitive web applications should always employ secure cookies, i.e., authentication cookies that are only transmitted over encrypted channels. However, this simple countermeasure is not always effective. For instance, in 2008 the Cookiemonster attack [24] highlighted vulnerabilities derived from an improper mixed support of secure and insecure connections. Cookiemonster captures cookies of improperly secured HTTPS sites via the local network. It tracks the HTTPS sites visited by a local client and automatically injects HTML elements for each HTTPS domain into subsequent regular HTTP requests. This causes any insecure HTTPS cookies from the target domains to be transmitted in clear and to be captured (and potentially replayed) by Cookiemonster. At the time it was proposed, this attack could be used to compromise Gmail accounts whose users did not set the "Always use HTTPS" option in their account. Such an option enforced HTTPS connections for all Gmail communications and prevent the attack. However, it took Google more than one year to completely fix this vulnerability [26] by setting HTTPS in Gmail by default. Such an attack—as well as simple session hijacking—could not be be used to hijack the Web History, but is an interesting example of vulnerabilities in web applications that do not *properly* provide mixed HTTP/HTTPS support.

**Privacy Threats.** Recent work has discussed potential privacy threats related to cloud service providers. For instance, [13] discussed potential threats and countermeasures associated with many forms of web activity—focusing on Google—related to the information collected by service providers. However, as opposed to our work, this paper focuses on the privacy threats against the service provider. Another direction was taken in [15,14] to assess user perception on alleged privacy threats by interviewing users. Among the other interesting results, it has been shown that more than 80% of users admitted to having conducted searches for information they would not want disclosed to their current or future employer.

**Limiting personal information disclosure.** Several techniques have been proposed to avoid user profiling and reduce the amount of information potentially leaked. For instance, the Firefox extension Trackmenot [19] periodically issues randomized search-queries to search engines to populate a user's search history with (non-clicked) queries and hide real queries. However, this would not prevent the Historiographer from retrieving "clicked" queries from the history and retrieve sensitive information.

**Auto-completion.** Finally, [10] mines Google's suggestion results to assess overall popular queries. However, the scope of the work is different, as they do not try to access single user's searches but popular ones using only generic suggestions.

**User categorization through search queries.** Several studies based on machine learning [23,27] use search histories to profile users and provide more accurate search results. This profile could be, for instance, a set of topics and categories of user's interest. Similar techniques could be used on search histories stolen by the Historiographer in order to categorize and profile victims.

# 7 Conclusion

This paper has presented a study of the private information disclosed to third parties from web searches. We showed that the well known session hijacking attack is still applicable to many Google services, and we presented the Historiographer, an attack that reconstructs Google's search histories from simple web searches. We have validated our technique through a large-scale experimental analysis.

We argue that solutions should be quickly deployed to protect users against these two types of attacks. The session hijacking attack is harmful not only because it allows an attacker to collect a lot of private information, including sometimes the search history, but also because it can be exploited to add potentially *compromising* entries [25]. It can also be used to modify the search results displayed to the victim. In fact, Google allows to delete or promote—i.e., show as first—results using a button associated to them. An adversary hijacking a session cookie can perform searches on the victim's behalf and influence the results corresponding to these searches as she wishes. For instance, this attack can be a powerful tool for censorship, as it can be used to remove or promote some pages displayed after a Google search.

The Historiographer can be used to reconstruct part of the Web History, when, for example, the simple session hijacking attack is not applicable. In addition, it can be used as an oracle to perform *targeted checks*, e.g., to verify the existence in the search history of specific keywords. The Historiographer is an *amplification* attack, and therefore is much more powerful than a simple eavesdropping attack: It not only allows an attacker to eavesdrop on the victim's search requests, but also allows him to retrieve the victim's *previous* search requests, *possibly performed from different networks and even different computers*. The number of potential victims is very high, since any signed-in user is at risk as soon as she issues a single Google search request from an unencrypted network, such as an open wireless network at an airport or a cafe.

These attacks deserve serious attention since Web Histories contain sensitive information. Any information leaked from Web search histories could endanger user privacy. Information retrieved from the search history could also be combined with other publicly available data, such as that published on social networks to accurately profile and/or identify target users. Furthermore, since the Historiographer also works for Google searches performed from mobile devices and such searches contain also localized results, one could use location-based services to also track users' movements and locations.

Although the Historiographer builds on features specific to the Google architecture, our goal is not to attack Google nor any particular service provider. Instead, we highlight the general problem of protecting the privacy of sensitive data when using a mixed architecture with both secure and insecure connections. As mentioned in [8], Google is not the only provider which leaves its customers vulnerable to data theft and account hijacking. As a matter of fact, the Bing search engine recently added a similar functionality to Personalized Suggestions. Users receive suggestions based on their previous searches and they can access the full search history [6]. Differently from Google, Bing only uses anonymous cookies for this purpose and stores the search history only up to 29 days. However, in Bing the full history is accessible via a simple session hijacking. We defer to future work a complete analysis of Bing and other search engines.

## A  Complete and Reconstructed History

Figure 6 and Example 2 show a portion of a complete Web History and the corresponding search history reconstructed using the Historiographer. Note that the complete Web History contains additional information for each query, which is not available in the reconstructed history. Reconstructed history entries only come from searches whose results have been clicked. In this example `pets 10` was not clicked and it does not appear in the reconstructed history.

```
pets 2010
privacy
privacy enhancing technologies symposium 2010
```

Example 2: Portion of the corresponding reconstructed history.

## B  Survey data

To complement our experimental analysis, we recruited 68 participants (36 males, 32 females) through various mailing-lists and social networking sites to complete an online survey. The results of our survey are shown in Table 4. Significant percentage of participants in our sample: (1) have a Google account (92.6%), (2) are likely to perform searches in Google while signed in (71.9%), since they do not normally sign out of Gmail, (3) have the Web History service enabled (60.6%). These high percentages in our sample suggests that the portion of Internet users that is vulnerable to the attack is quite significant.

Fig. 6: Portion of a complete web history retrieved from one of the authors' account.

Additionally, 53.7% of users were concerned about the fact that Google keeps track of their web searches and 65.2% considered their searches as sensitive information. Finally, a large majority of users were concerned by potential disclosure of web searches: 76.5% declared to have conducted a web search related to information they were reluctant to disclose to an employer, and 70.3% to anyone. Note that this results are similar to a study performed in 2007 [15].

| USERS | % |
| --- | --- |
| With a Google account | 92.6% |
| Not signing out of Gmail | 71.9% |
| With Web History enabled | 60.6% |
| Concerned with Google tracking their searches | 53.7% |
| Considering their searches sensitive information | 60.6% |
| Conducted a search not to be disclosed to employer | 76.5% |
| Conducted a search not to be disclosed to anyone | 70.3% |

Table 4: Results from the survey.